\documentclass[12pt]{iopart}

\usepackage{graphicx}

\begin{document}

\title{Fractal dimensions of the $Q$-state Potts model for the complete and external hulls}

\author{David A. Adams$^1$,  Leonard M. Sander$^2$, and Robert M. Ziff$^3$
}


\address{$^1$ Department of Physics, University of Michigan, Ann Arbor MI, 48109, USA \\
             $^2$ Department of Physics and Michigan Center for Theoretical Physics, University of Michigan, Ann Arbor MI, 48109, USA \\
              $^3$ Department of Chemical Engineering and Michigan Center for Theoretical Physics, University of Michigan, Ann Arbor MI,48109, USA \\
Email: davidada@umich.edu, lsander@umich.edu, rziff@umich.edu 
}

\begin{abstract}
Fortuin-Kastelyn clusters in the critical $Q$-state Potts model are conformally invariant fractals. We obtain simulation results for the fractal dimension of the complete and external (accessible) hulls for $Q=1$, $2$, $3$, and $4$, on clusters that wrap around a cylindrical system.  We find excellent agreement between these results and theoretical predictions.  We also obtain the probability distributions of the hull lengths and maximal heights of the clusters in this geometry and provide a conjecture for their form.
\end{abstract}

\vspace{2pc}
\maketitle

\section{Introduction}

The $Q$-state Potts model \cite{Potts52} is a well-studied system in condensed matter physics, exhibiting a continuous phase transition for $Q \le 4$ \cite{Wu82}. \   It is a generalization of the Ising model \cite{Ising25} with $Q$ different spins, such that like spins interact with a single coupling constant $K$, and has applications  to a range of different physical systems:  $Q=1$ and $Q=2$ correspond to percolation and the Ising model, respectively, and   $Q=3$ has been used to represent absorbed rare-gas monolayers on graphite surfaces \cite{Alexander75,Bretz77}. \  In this paper we present results from a numerical study of the fractal dimensions of the hulls of the Fortuin-Kastelyn (FK) clusters \cite{KF69,FK72} for $Q=1,2,3,4$ using a cylindrical system geometry. In addition we look at the probability distribution of the lengths of these hulls, and find that it has a simple exponential tail.  A related quantity, the probability distribution of the height (i.e. the vertical span) of the hulls, also has such a tail. 

Fractal clusters enter the Potts model in the following way: for percolation, $Q=1$, it is well-known that the spanning cluster at the critical point is a fractal. The generalization of this for other $Q$ is that at the critical point, $K_c$,  a  subset of clusters of like spins, the FK clusters \cite{KF69,FK72},  are  fractal \cite{Voss84}. \  FK introduced the clusters by showing that the partition function of the $Q$-state Potts model can be written as a sum over all bond configurations on a lattice  with a bond occupation probability $p = 1 - e^{-K}$, multiplied by a weight $Q^{N_c}$ where $N_c$ is the number of clusters in a distinct configuration.  To go from a spin configuration to a corresponding bond configuration, bonds are added between like spins with probability $p$; the connected clusters are the ones we want. The partition function for the Potts model is thus a sum over all possible FK clusters, and at the critical point $p=p_c=1-e^{-K_c}$, the clusters are fractal. The  $p_c(Q)$  for the triangular lattice are given by  \cite{Kim74}:
\begin{eqnarray}
Q=2: &\quad& p_c(2) = 1 - 1/\sqrt{3} \approx 0.42265, \nonumber \\
Q=3: &\quad& p_c(3) = 1 -\left[1+ \frac{1}{2}\sqrt{3}\sec{\left(\frac{\pi}{18}\right)}\right]^{-1} \approx 0.46791, \nonumber \\
Q=4: &\quad& p_c(4) = 1/2. \nonumber \\
\end{eqnarray}

These clusters are characterized by several different fractal dimensions, including $D_H$, $D_{EP}$, $D_M$, $D_{SC}$, $D_G$, corresponding to the complete hull, external hull, mass, singly connected bonds, and narrow-gate fjords.  Most  previous work on the fractal properties of these clusters has focused on percolation ($Q=1$) for which numerical values have been given for $D_H$ \cite{Voss84,Ziff84,Grassberger86,SapovalRossoGouyet85}, $D_{EP}$ \cite{Grossman86}, $D_{SC}$ \cite{Pike81}, and $D_G$ \cite{Asikainen03}. \   Theoretical studies for percolation include studies of $D_{SC}$ \cite{Coniglio81,Coniglio82}, $D_H$ \cite{Bunde85}, $D_{EP}$ \cite{Grossman86}  and $D_G$ \cite{Aizenman99}.   \   Theoretical values for $D_M$, $D_H$, and $D_{SC}$ have been calculated for all $Q$ values up to the upper critical dimension $Q=4$ \cite{Saleur87,Duplantier99,Duplantier00}. \     Hull exponents have also been derived by SLE theory \cite{SmirnovWerner01}. 

In this paper will present numerical data for $D_H$ and $D_{EP}$ and the height and length distributions for both types of perimeters. In a recent paper, Asikainen et al.\ \cite{Asikainen03} measured $D_M$, $D_H$, $D_{EP}$, $D_{SC}$, $D_G$ for $Q=1$ through $Q=4$  by studying individual isolated clusters that do not touch the boundary.  In the present work we consider a cylindrical geometry and look at the hulls of the clusters that wrap around it.  This method provides an unambiguous measure of the length scale (namely, the circumference) and leads to accurate numerical results for the fractal dimensions.  We also measure the height and length distributions of the hulls themselves.
\section{Model}

Two types of simulations are used.  To model $Q=1$ (percolation), we use the Leath algorithm \cite{Leath} with site percolation. \   This method starts with a single ``active" site, which attempts to change its undetermined neighbors into active sites with a probability $p_c$. \   If a change fails, the site is becomes ``inactive", and can never be made active.  Each new active site attempts to make all of its neighbors active until no more active sites remain.  The resulting structure is a percolation cluster. All clusters are grown on the triangular lattice.

For all other $Q$-states, we generate FK clusters \cite{KF69,FK72} using the Swendsen-Wang (SW) method \cite{Swendsen87}. \   
The SW algorithm is as follows: after the bonds have been placed for a given configuration, each cluster of sites connected by bonds is labeled with a randomly chosen spin.   Then new bonds are put down with probability $p$ between like spins, forming the FK clusters.  This process is then repeated.  Each cycle constitutes a spin update in that we have flipped entire FK clusters. The SW method allows for fast equilibration of critical clusters.

\section{Simulation}

We grew our clusters on a triangular lattice with periodic boundary conditions along both directions.  We made the dimensions of system elongated so that most large clusters wrapped around in one direction but not the other. The system, though really a torus, was effectively a cylinder.  Clusters that wrapped around both directions were rejected.   The lattices had an aspect ratio of $100$:$1$ and $8$:$1$ for percolation and FK clusters, respectively. We use the width ($W$), of the lattice as the characteristic length.  For $Q=1$, clusters were grown for $W=8$, $16$, $32$, $64$, $128$, and $256$. \  For $Q=2$ through $4$, we also considered $W = 512$ and $1024$ in order to reach the asymptotic fractal dimension.

For each valid cluster grown, the number of sites on the top and bottom of the cluster were recorded as well as the height (vertical span) of the hulls, defined as the highest point on the top hull minus the lowest point on the top hull. The top and bottom hulls were considered independently and both were used to calculate the fractal dimension.  To record the complete hull, we used the outermost layer of active sites as the hull.  For the external hull, we used the layer of inactive sites that pad (are neighbors to) the outermost layer of active sites.

\begin{table}
\caption{\label{fig:Tbl1} The theoretical \cite{Saleur87,Duplantier99,Duplantier00} and measured values for the fractal dimensions for the complete (C) and external (E) hulls of $Q$-state Potts model}
\begin{tabular}{|c | c | c |} \hline
 $Q$ & theory & measured \\ \hline
 1C & 7/4 = 1.750 & 1.747 \\ \hline
 1E & 4/3 = 1.333 & 1.330 \\ \hline
 2C & 5/3 = 1.667 & 1.663 \\ \hline
 2E & 11/8 = 1.375 & 1.375 \\ \hline
 3C & 8/5 = 1.600 & 1.602 \\ \hline
 3E & 17/12 = 1.417 & 1.412 \\ \hline
 4C & 3/2 = 1.500& 1.510 \\ \hline
 4E & 3/2 = 1.500& 1.534 \\ \hline
\end{tabular}
\end{table}

In the case of FK clusters, the system must be equilibrated before recording can start.  The equilibration times used are $2W$, $3W$, and $16W$, for $Q=2$, $3$, and $4$, respectively.  We use longer equilibrium times for $Q=4$ because of its slower dynamics \cite{LiSokal89}. \   To determine the equilibration time, we measure the average energy per spin and the average largest cluster size as a function of the number of spin updates. We find that both obserables relax to their steady-state values exponentially quickly with the same decay time for a given $Q$ and $W$, though the average energy is always closer to the steady-state value at a given step. For example, we find the average energy was within $1$\% of the steady-state value within $500$ spin updates, whereas it took the average largest cluster size $1500$ spin updates to reach $1$\% of its steady-state value for $Q=4$ and $W=256$. Because of its slower relaxation, we believe that the average largest cluster size is a better measure of equilibration.

After the system is equilibrated, we pick a random cluster, and if the cluster wraps around the width of the system, its hull length and height are recorded in a similar fashion as with percolation.  Note that while the FK clusters are essentially bond percolation clusters, we consider the hulls on the sites of the clusters -- that is, we treat the sites in the FK clusters as a site percolation problem. To record the external hulls of these clusters a layer of inactive sites are simply added outside of the top and bottom hulls.  (This is the advantage of using the triangular lattice.)  Several spin updates are taken between each successive attempt to find a valid cluster.  The number of updates between measurements are $8$, $12$, and $64$ for $Q=2$, $3$, and $4$, respectively.

The work of Asikainen et al.\ \cite{Asikainen03} differs with ours in several ways.  For $Q>1$, they simulated a single system size, $4096 \times 4096$, and picked random clusters after the system was well equilibrated. To equilibrate such a large system, they needed to use a new technique to speed up equilibration. For their system of isolated clusters, the length scale is not as clearly
defined as it is in the cylindrical geometry; they used the cluster's radius
of gyration for the length scale. Lastly, they used a square lattice for their simulations, for which the complete and external hulls are not as clearly defined as in the triangular case because of the corners in the square system \cite{Grossman87}. \  Our results are consistent with theirs in regards to the fractal dimensions, but are significantly more precise.

\section{Results}

For each state $Q$ we obtain a large set of values for the lengths of the complete and external hulls and hull heights for wide range of system widths.  With this data we can compute the fractal dimension of the different hulls.  For a fractal the length of the hulls scales as a power $L \propto W^{D}$, where $D$ is the fractal dimension.  We measure $D$ in the conventional way  by making a linear fit  of $\ln L$ as a function of $\ln W$. \   The accuracy can be assessed two ways: (1) the value of the square of the Pearson product-moment coefficient, $R^2$, of the fit and (2) the apparent randomness of the residuals.  We focus on the latter as it can be used to determine when finite-size effects are significant, and thus which sizes we can use for the fit.  We determined that all measured widths for the complete hull can be used to fit $D$ for $Q<4$. \   For the external hull, there are significant corrections for $W < 256$, so for those systems, we use $256$, $512$, and $1024$. \   We use the same three sizes for both hulls of $Q=4$. \ 

\begin{figure}
\includegraphics[width=0.75\textwidth]{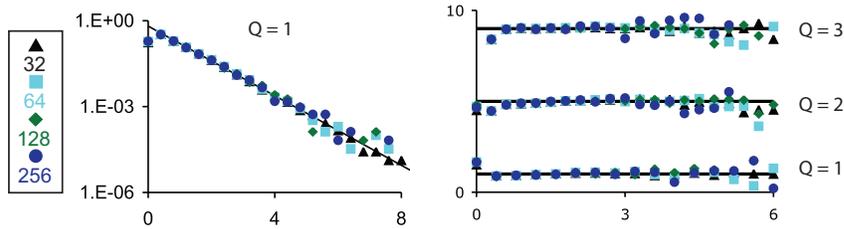}
\caption{\label{fig:ProbDistQC} The probability distributions for the lengths of the complete hulls, for several system widths: $W=32$, $64$, $128$ and $256$ for $Q=1$ as a function of $L/\langle L \rangle$ where $\langle L \rangle = 0.93 W^{7/4}$ (left). The data are put into bins of size $C \langle L \rangle$ with $C = 0.4$. \   The black line is the best fit using an exponential with an inverse decay length of $\lambda^{-1}$. \  The plot on the right is of the residual (difference) of the fit of exponentials to the different $Q$ complete hulls, where different $C =0.3$ for $Q = 2, 3$. The best fit for the $\lambda^{-1}$ for $Q=1$, $2$, and $3$ are $1.4$, $1.65$, and $1.7$, respectively.
 The residual data for different $Q$ are offset vertically for clarity.  }
\end{figure}

Table \ref{fig:Tbl1} shows a summary of our findings for the fractal dimensions of the complete and external hulls for $Q=1,2,3,4$. \   With the exception of the external hull for $Q=4$, all the measured values agree with theory within a fraction of a percent.  The discrepancy at $Q=4$  can be interpreted as arising from logarithmic corrections to scaling \cite{Aharony03}. \   In Ref \cite{Asikainen03} a correction is made for the slow crossover. 

\begin{figure}
\includegraphics[width=0.75\textwidth]{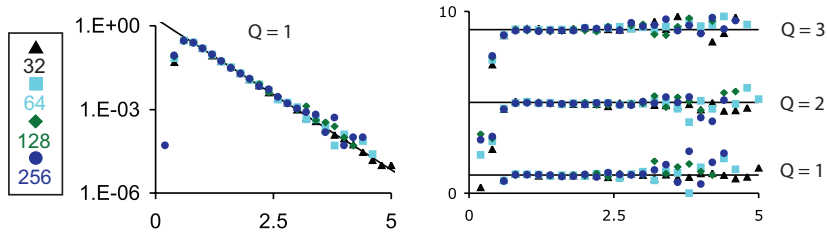}
\caption{\label{fig:ProbDistQE} The probability distributions for the lengths of the external hulls, for several system widths: $W=32$, $W=64$, $W=128$, $W=256$ for $Q=1$ (left) where the different system sizes are scaled in the same way as fig.~\ref{fig:ProbDistQC}. The right shows the residuals for $Q=1$ through $Q=3$. \  The best fit for $\lambda^{-1}$ for the different $Q$'s are $2.5$, $2.4$, and $2.35$ for $Q=1$, $2$, and $3$, respectively. The residual data for different $Q$ are offset vertically for clarity.}
\end{figure}

We now turn to the probability distribution of the quantities that we measured. We first found that  the average height of the hulls scales linearly with system width so that these are isotropic fractals.  
We then measured the heights and hull lengths of a large number of clusters in order  to produce the probability distribution of hull lengths.  Figure \ref{fig:ProbDistQC} shows the scaled probability distributions of complete hulls for $Q=1$ and the residuals to the best-fit exponential to the first three values of $Q$. \   $Q=4$ is not shown because the scaling does not work for the size systems we used.  Figure \ref{fig:ProbDistQE} is a similar plot for the external hulls, with a finer bin size.  All scaled hull-length distributions have exponential tails. The complete and external hulls appear to have fairly similar decay lengths for the different values of $Q$.

\begin{figure}
\includegraphics[width=0.75\textwidth]{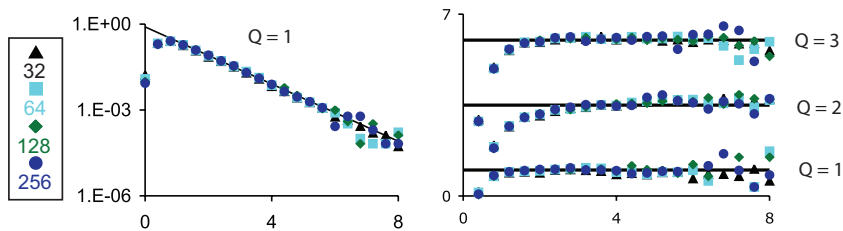}
\caption{\label{fig:ProbDistCHeight} The probability distributions for the heights of the complete hulls, for several system widths: $W=32$, $W=64$, $W=128$, $W=256$ for $Q=1$ (left).  The bin sizes are $CW$, where $C = 0.4$ . The right shows the residuals for $Q=1$ through $Q=3$. \  The best fit for $\lambda^{-1}$ for the different $Q$'s are $1.15$, $1.0$, and $1.05$ for $Q=1$, $2$, and $3$, respectively. The residual data for different $Q$ are offset vertically for clarity.}
\end{figure}

\begin{figure}
\includegraphics[width=0.75\textwidth]{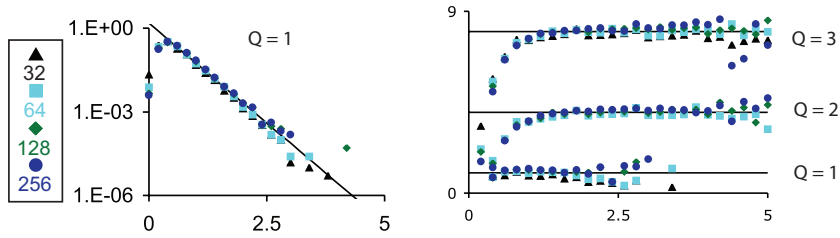}
\caption{\label{fig:ProbDistEHeight} The probability distributions for the heights of the external hulls, for several system widths: $W=32$, $W=64$, $W=128$, $W=256$ for $Q=1$ (left). The bin sizes are $CW$, where $C = 0.2$ . The right shows the residuals for $Q=1$ through $Q=3$. \  The best fit for $\lambda^{-1}$ for the different $Q$'s are $3.3$, $1.8$, and $1.7$ for $Q=1$, $2$, and $3$, respectively. The residual data for different $Q$ are offset vertically for clarity.}
\end{figure}

We also calculate the probability distribution of hull heights for the complete and external hulls, Figures \ref{fig:ProbDistCHeight} and \ref{fig:ProbDistEHeight}. \    Again we see that the distributions have exponential tails.

The exponential tails are, in some sense, no surprise in this type of problem as we can see from a  simple example. Consider the track of a free random walker in a channel with absorbing walls at $x=\pm W/2$. \  The walker starts in the middle of the channel, and we seek the probability distribution of the maximum height attained by the walk before it hits the side walls. This is more-or-less what we are doing with our spanning fractals, though they are not free random walks, of course.

It is not hard to see why we get exponential tails in this case, by considering the following steps: first we define an auxiliary problem by putting an absorbing wall at height $h$ above the origin. The number of walks, $N(h)$, that ever hit this wall before being absorbed on the sides is the number of walks with height greater than $h$. \  The distribution we seek is proportional to $dN/dh$. 
The problem can be solved exactly by going to the continuum limit and using conformal mapping. However, we can guess the solution easily: the probability for a walker to penetrate a channel for height $h$ is clearly exponentially decreasing in $h$; the conformal map gives $e^{-\pi h/W}$. \  Then $N$ must have this dependence, along with its derivative.  

For percolation or FK clusters, the probability distribution of the maximum height can also be found
from the derivative of the probability of  crossing, and there one also
expects exponential behavior for large $h$ \cite{Cardy92,Cardy06}.  

The existence of exponential tails in both the hull length and height distributions may appear contradictory.  We know that the average height, $h$, is proportional to the system width, $W$ which is related to hull length, $L$ through a power-law $W \sim L^{1/D_H}$.  One might guess that this  should lead to a stretched exponential $\exp(-cL^{1/D_H})$ rather than simple exponential for the distribution of $L$. \   In order to clarify this issue, we looked at another well-studied system: the percolation hull-walk inÊ an open rectangular region \cite{Ziff84} for which it is easy to generate a large ensemble of fractal curves.   

We performed simulations of the self-avoiding hull walker on a square lattice
\cite{Ziff84,Grassberger86}. The hull walk algorithm is as follows.  Every step the walker either turns right or left.  If the current site hasn't been marked, the walker turns left with probability $p$, marks the site `active' and takes a step forward.  With probability $1-p$ it turns right, marks the site inactive, and takes a step forward.  If the walker steps onto a active or inactive site it always turns left or right, respectively.  We used this walk  to create closed figures that have the same fractal dimension as the complete hull of percolation clusters, $D=7/4$. \   We recorded the average walk length as a function of the maximum height for several values of $W$. \  

For every W, the walk length is proportional to the maximum height,Ê
i.e., $L \sim m_W h$, as one would expect for large $h$. ÊHowever, we find
that the \emph{coefficient}, $m_W$, depends on $W$ as $m_WÊ= a W^{3/4}$.    With these pieces, we get back the known power-law relationship between $L$ and $W$, $\langle L \rangle \sim m_W \langle h \rangle \approx a W^{3/4} W = a W^{7/4} $, so that we recover the fractal dimension of $7/4$. \  

\section{Conclusion}

In this paper, we measured the fractal dimensions for the complete and external hull lengths, $D_H$ and $D_{EP}$ for $Q=1$ through $Q=4$. We used the Leath algorithm to grow critical percolation clusters. For $Q=2,3,4$, we used the SW method to generate critical FK clusters.  All systems used a triangular lattice with periodic boundary conditions.  The aspect ratio of the systems were heavily skewed so that spanning clusters would span and one direction and not the other. The smaller length (W) exactly determined the length scale of the system. We generated a large number of spanning clusters of various system sizes,  $W = 8,16,32,64,128,256$ and additionally $W=512,1024$, for percolation and $Q > 1$, respectively. We find excellent agreement between our results and the associated theories \cite{Saleur87,Duplantier99,Duplantier00}. \   We also measured the distributions of the hull lengths and heights.  We found that the distributions for $Q=1,2,3$ have exponential tails.  The values of the inverse decay lengths for the exponentials are given in the figure captions.We discussed the apparent contradiction between the height and length distributions both having exponential tails, as opposed to one being a stretched exponential.  We resolved this contradiction by showed that the relationship between average height and width is in fact linear for a fixed $W$. \   We found that the power-law scaling formula can be written as $\langle L \rangle \sim m_W \langle H \rangle \approx a W^{3/4} W $. \   This formula clarifies the relationships between $L$, $H$, and $W$. \ We are not aware of any predictions regarding the exponential tails in the hull length distributions.

\section*{Acknowledgments}This work was supported in part by the National Science Foundation through DMS-0553487. We would like to thank C. Doering for a useful discussion.

\section*{References}
\bibliographystyle{unsrt} 
\bibliography{Draft_Fractal_Bib}

\end{document}